\documentclass[10pt]{iopart}
\begin{document}
\title[ Hamiltonian Frenet-Serret dynamics]{Hamiltonian 
Frenet-Serret dynamics}

\author{R Capovilla${}^{\dagger}$ ,
J Guven${}^\ddagger$  and 
E Rojas${}^{\dagger\S}$}
\address{$\dagger$\
Departamento de F\'{\i}sica,
Centro de Investigaci\'on y de Estudios Avanzados del IPN, 
 Apdo. Postal 14-740,07000 M\'exico, DF, MEXICO}
\address{ $\ddagger$\ Instituto de Ciencias Nucleares,
Universidad Nacional Aut\'onoma de M\'exico,
 Apdo. Postal 70-543, 04510 M\'exico, DF, MEXICO}
\address{ $\S$\ Department of Physics,
Syracuse University,
Syracuse, NY 13244-1130, USA}

\begin{abstract}
The Hamiltonian formulation of the dynamics of a 
relativistic particle described
by a higher-derivative action that depends 
both on the first and the second
Frenet-Serret curvatures is considered from 
a geometrical perspective.
We demonstrate how 
reparametrization covariant dynamical 
variables and their projections onto
the Frenet-Serret frame can be exploited 
to provide not only a 
significant simplification of but also 
novel insights into 
the canonical analysis. 
The constraint algebra and the Hamiltonian 
equations of motion
are written down and a geometrical 
interpretation is provided
for the canonical variables.
\end{abstract}



\section{Introduction}

Geometrical models that describe
a relativistic particle may be constructed
using the
geometrical scalars associated with the
embedding of the particle worldline in spacetime
as building blocks for the action.
The simplest model of this kind is
the massive free particle, described by
an action which is proportional to the proper
time along the worldline. In fact, in the absence of
external fields, this is also
the only reparametrization invariant
action one can
write down solely in terms of the velocity.
The next simplest model involves the curvature
of the worldline, and  is
acceleration dependent. 

The study of such models was initially
motivated by the suggestion by Polyakov \cite{Polyakov} (and
independently by Kleinert \cite{Kleinert}), of modifiying
the Nambu-Goto action for a relativistic string model
of QCD, by adding a term quadratic in the mean extrinsic
curvature of the worldsheet which is scale invariant.
Pisarski took the natural first step
to try to understand its scale invariant
point-like analogue \cite{Pisarski}.
Since then, the exploration of these geometrical
particle models has taken on a life of
its own, with a case by case analysis
of the classical dynamics as well as
peculiarities related to their quantization
pioneered by Plyushchay \cite{Ply1,Ply2,Ply3,Ply4,Ply5}
(see also \cite{Nes1,Nes2,Nes3,Nerse}). In other fields, like
in $2 + 1$-dimensional gauge field theories, systems of
relativistic particles involving curvature of the worldline
as well as its torsion, have been considered as models for a
massive relativistic anyon \cite{KP}.

We consider  the Hamiltonian
formulation for geometrical models,
describing a relativistic point-like object moving
in Minkowski spacetime.
With respect to previous approaches, however, we
focus on the common features of models described by an action that
depends in an arbitrary way on the first and second Frenet-Serret 
[F-S] curvatures of the worldline.
We are more interested in the general structure of the
Hamiltonian formulation of
reparametrization invariant higher-order constrained
systems than in the specific details of any particular model
with an eye towards possible extensions to strings
and branes.
We adopt a geometrical language
for the worldline. We use a set of reparametrization
covariant variables for the configuration space.
This step alone produces unexpected simplifications
in the canonical analysis, compared to the obvious choice of the embedding
functions and their derivatives with respect to an
arbitrary parameter along the worldline as configuration
variables. The natural adapted frame to the particle worldline for 
higher derivative theories is the F-S frame. 
By a judicious use of the F-S equations in the
intermediate steps of the canonical analysis, we are able to treat
in full  generality models that depend on the first or on the
second F-S curvature.   
We abstain from the practice which is standard of 
introducing additional constraints and auxiliary variables in order 
to cast the model in terms of velocities alone. Such a strategy, in our view,
although useful in special cases, only serves to obscure the 
elegant canonical structure
possessed by these models.

This paper provides a companion to \cite{ACG2} where  
the corresponding Lagrangian analysis is undertaken.

The paper is organized as follows.
In Sect. 2, we briefly recall some basic facts about the geometry of
a timelike curve in Minkowski spacetime \cite{ACG2},
and we collect some useful
formulae, to be used later in the Hamiltonian analysis. 
In Sect. 3, we describe the Hamiltonian
formulation for higher-derivative theories, using reparametrization
covariant variables, and how it is to be adapted when constraints arise,
as in our case, corresponding to reparametrization invariance. 

In Sect. 4,
we consider a geometrical model for a relativistic
particle that depends at most on second derivatives
of the embedding functions. The most general Lagrangian
which satisfies the requirements of reparametrization invariance
and invariance under rotations of the normals to the
worldline is an arbitrary function of the first F-S curvature.
We derive the form of the momenta in the extended phase space
appropriate for a higher-derivative theory, and provide their
geometrical interpretation in terms of the F-S basis. In particular we find
that, up to a model dependent proportionality factor,
the velocity is conjugate to the first F-S normal to the worldline.
As expected
from reparametrization invariance, there is one primary constraint
and its conservation in time implies that the canonical Hamiltonian
vanishes.
We compare the  Hamilton equations with the corresponding Lagrangian equations  of
motion. For illustration purposes, we consider a  representative example:
 a model quadratic in the geodesic curvature.

In Sect. 5, we extend our considerations to models
that depend arbitrarily on the second F-S curvature
of the worldline.
We again obtain explicit expressions for the
momenta. We uncover a surprising analogy between this class of models
and the ones considered in Sect. 4. Here the momenta conjugate to the
acceleration is proportional to the second F-S normal. There are two
primary constraints. Since only first class primary constraints 
correspond to symmetries of the action, one of the two has to be
second class, as we verify explicitly. The complete constraint algebra
as well as the Hamiltonian equations of motion are written down. 

We conclude with a few general remarks in Sect. 6.

\section{Worldline geometry}

We begin by recalling  some basic facts about the geometry of
a timelike worldline in Minkowski spacetime, in order to establish
our notation. We also collect some useful expressions that will be
used below, in the Hamiltonian analysis.

We consider a relativistic particle with worldline
described by the embedding functions  $ x^\mu = X^\mu (\xi)$, where
$x^\mu$ are local coordinates for the background
Minkowski spacetime
$(\mu,\nu,\dots = 0,1,\cdots, N)$, and $\xi$ is an arbitrary parameter.

The vector tangent to the worldline is
$ \dot{X}^\mu = d X^\mu / d\xi $,
and the one-dimensional metric along the worldline is
$\gamma = \dot{X}^\mu \dot{X}^\nu \eta_{\mu\nu} = \dot{X}\cdot
\dot{X} $, where $\eta_{\mu\nu}$
is the Minkowski metric
with only one minus sign.
We assume that the
worldline is timelike,  $ \dot{X}^2 < 0$.
The proper time along the worldline is given
infinitesimally by $d\tau = \sqrt{-\gamma} \; d\xi$.
We use a prime to denote differentiation by proper
time. Taking advantage of the
fact that the intrinsic geometry of a curve is trivial
in the parametrization
by $\tau$, we have $\gamma = X' \cdot X' = -1$. 
The affine connection along the curve is
$\Gamma = \gamma^{-1} \dot{X} \cdot \ddot{X}$.
We use it define the covariant derivative under
reparametrizations $\nabla = (d/ d\xi ) - \Gamma $.
Note that $\nabla X = \dot{X} $ and that $\dot{X} \cdot
\nabla^2 X = 0$. $\Gamma$ vanishes in a parametrization by proper time.

The extrinsic curvature along the $i-$th normal is defined by
$K^i = - n^i \cdot \ddot{X} = - n^i \cdot \nabla^2 X$,
where the $i$-th normal to the worldline is
defined implicitly by
$n^i \cdot \dot{X}  = 0 $,
and we choose to normalize them
$n^i \cdot n^j  = \delta^{ij}$ ($i,j \dots = 1,2,\cdots, N$).
$K^i$ is the point-like analogue
of the extrinsic curvature for higher dimensional objects.
The point-like analogue of the mean extrinsic
curvature is
\begin{equation}
k^i = \gamma^{-1} K^i = (-\gamma)^{-1}
n^i \cdot \nabla^2 X\,.
\end{equation}
This quantity
 transforms
as a scalar under worldline reparametrizations, whereas
$K^i$ itself does not.
The geodesic curvature is given by the magnitude of $k^i$,
\begin{equation}
 k = \sqrt{ k^i \; k_i }\,.
\label{eq:k1}
\end{equation}
In addition the extrinsic geometry of the worldline is described
by a connection associated with the freedom to rotate the normal
vectors, $\omega^{ij} = \dot{n}^i \cdot n^j$. We use it to 
define a derivative also covariant under rotations,
$\widetilde\nabla = (d / d\xi ) - \Gamma - \omega$.

An alternative description of the worldline geometry
is given in terms of the orthonormal F-S basis
$\{ X' , \eta_i \}$.  The F-S equations
for a curve in $(N+1)$-dimensional Minkowski spacetime  are
\cite{ACG2,Spivak}
\begin{eqnarray}
X'' &=& \kappa_1 \; \eta_1 \,,
\nonumber
\\
\eta_1{}' &=& \kappa_1 \; X' - \kappa_2 \; \eta_2 \,,
\nonumber \\
\eta_2{}' &=& \kappa_2 \; \eta_1 - \kappa_3 \; \eta_3 \,,
\nonumber \\
\dots      &{}& \dots  \label{eq:frenet} \\
\eta_{N-1}{}' &=& \kappa_{N-1} \; \eta_{N-2} - \kappa_N \; \eta_N \,,
\nonumber\\
\eta_N{}' &=& \kappa_N \; \eta_{N-1} \,.
\nonumber
\end{eqnarray}
where $\kappa_i $ denotes the $i$-th Frenet-Serret curvature.
$\kappa_1 $ is the geodesic curvature, $k$.
In
an ambient spacetime of dimension $N+1$ there are at most $N$ F-S
curvatures. 
Implicit in the F-S construction is that the
embedding functions are $N+1$ times differentiable, and that the
$\kappa_i$ may vanish only at isolated points \cite{Spivak}.
Indeed, if $\kappa_i$ vanishes so do all the higher ones and the
worldline lies in a $i+1$-dimensional subspace. An important
result of the geometry of curves is that the curvatures determine
the embedding functions by quadratures, up to rigid Poincar\'e
motions \cite{Spivak}. Therefore the curvatures can be used as a natural
set of
variables in the description of the worldline.

The relationship between these two descriptions of the worldline
geometry is spelled out {\it e.g.} in  Ref. \cite{ACG2}. We recall
here only some specific expressions that will be needed in this
paper.
 In terms of (derivatives of) $k^i$, the second
F-S curvature can be written as
\begin{equation}
\kappa_2 = {1 \over \sqrt{-\gamma}} \; 
\sqrt{\widetilde\nabla \hat{k}^i \; \widetilde\nabla
\hat{k}_i }\,,
\label{eq:kappa2}
\end{equation}
where $\hat{k}^i = k^i / k$. The relation between the first two
F-S normals and $k^i$ and its derivatives is given by
\begin{eqnarray}
\eta_1 &=& \hat{k}^i \; n_i = {1 \over (-\gamma) \kappa_1 } \nabla^2 X\,,
\label{eq:eta1} \\
\eta_2 &=& - {1 \over \kappa_2 \sqrt{-\gamma}} \;
\left( \tilde{\nabla} \hat{k}^i \right) \; 
n_i \,.
\label{eq:eta2}
\end{eqnarray}
Note  that the orthogonality $\eta_1 \cdot \eta_2
= 0 $ follows from the unit vector fact $\hat{k}_i
\widetilde\nabla \hat{k}^i = 0$.

For the purpose of performing a canonical analysis, we need
to express both $\kappa_1$ and $\kappa_2$ directly in terms of
derivatives of the embedding functions. We have
\begin{equation}
\kappa_1 = {1 \over (-\gamma)}
\sqrt{ ( n^i \cdot \nabla^2 X )\;  ( n_i \cdot \nabla^2 X )}
= {1 \over (- \gamma)} \sqrt{ \nabla^2 X
\cdot \nabla^2 X}\,,
\label{eq:k1d}
\end{equation}
where to obtain the second equality we have used the
completeness relation $\eta^{\mu\nu}
= \gamma^{-1} \dot{X}^\mu \dot{X}^\nu + n^{\mu \, i }
n^{\nu}{}_i $, and the orthogonality $\dot{X} \cdot \nabla^2 X
= 0$. For $\kappa_2$, note that
\begin{equation}
\widetilde\nabla k^i = (-\gamma )^{-1} [ (\widetilde\nabla n^i )
\cdot \nabla^2 X + n^i \cdot \nabla^3 X ] = (-\gamma )^{-1} n^i
\cdot \nabla^3 X\,.
\label{eq:eta2a}
\end{equation}
The first term in the first equality vanishes since
$\widetilde\nabla n^i = k^i \dot{X} $ and $\dot{X} \cdot \nabla^2
X = 0$. For the derivative of the unit vector $\hat{k}^i$ which
appears both in $\kappa_2$ and $\eta_2$, this implies
\begin{equation}
\hspace{-1.5cm}
(\tilde{\nabla} \hat{k}^i )\, n_i
= {1 \over \sqrt{ \nabla^2 X
\cdot \nabla^2 X } } \left[
\nabla^3 X
- { (\nabla^2 X \cdot \nabla^3 X ) 
\over (\nabla^2 X \cdot \nabla^2 X ) } \nabla^2 X
- {( \dot{X} \cdot \nabla^3 X ) \over (  \dot{X} \cdot \dot{X} ) }  
\dot{X} \right]\,.
\label{eq:eta2d}
\end{equation}
Note that we have the expected orthogonality conditions
$\tilde{\nabla} \hat{k}^i n_i \cdot \dot{X} = 0 , \tilde{\nabla}
\hat{k}^i n_i \cdot \nabla^2 X = 0$. Using now Eqs.
(\ref{eq:kappa2}), (\ref{eq:eta2d}), we obtain for $\kappa_2$ the
expression,
\begin{equation}
\hspace{-2cm}
\kappa_2 = {1 \over \sqrt{-\gamma}
\sqrt{ \nabla^2 X \cdot \nabla^2 X } } \left[
( \nabla^3 X  \cdot \nabla^3 X )
- { (\nabla^2 X \cdot \nabla^3 X )^2 \over (\nabla^2 X \cdot
\nabla^2 X )}
- {( \dot{X}\cdot \nabla^3 X )^2 \over (  \dot{X}\cdot \dot{X} ) }  
\right]^{1/2}.
\label{eq:k2d}
\end{equation}

\section{Hamiltonian formulation}

We consider a relativistic particle whose dynamics is
described by a local action invariant under Poincar\'{e}
transformations, under
worldline reparameterizations, and when the co-dimension
of the worldline is greater than one, under rotations
of the normals.
We specialize our considerations to geometric models
which depend at most on three derivatives of the
embedding functions $X$. Under these assumptions
the most general action is of the form
\begin{equation}
S = \int d\xi \; {\cal L} = \int d\xi \; \sqrt{-\gamma} \; 
L ( \kappa_1 , \kappa_2 , \kappa_1{}' )\,,
\end{equation}
where ${\cal L}$ is the Lagrangian density, and $L$ the
Lagrangian function. We will neglect the possibility
of a dependance on the derivative of $\kappa_1$ and
consider $L = L (\kappa_1 , \kappa_2 )$.

We are considering a higher-derivative model,
with $L = L ( \dot{X} , \ddot{X} , {\mathop{X}\limits^{\ldots}} )$. 
Note that, in
the absence of external fields, a dependance on the
embedding functions $X$ would break Poincar\'{e}
invariance. Rather than the obvious choice of
$( \dot{X} , \ddot{X} , {\mathop{X}\limits^{\ldots}} )$
as configuration variables, it turns out to be extremely
convenient to use the combinations covariant under
reparametrizations given by $ ( \dot{X}, \nabla^2 X ,
\nabla^3 X )$. Therefore we consider as phase space for these models
the conjugate pairs
$\{ p , X \; ; P , \dot{X} \; ; \Pi , \nabla^2 X \}$.
The momenta conjugated to
$\{ \nabla^2 X , \dot{X}, X \} $ are, respectively,
\begin{eqnarray}
\Pi &=& {\partial {\cal L} \over \partial \nabla^3 X }\,,
\label{eq:p1}
\\
P &=& {\partial {\cal L} \over \partial \nabla^2 X } -
\nabla \left( {\partial {\cal L} \over \partial \nabla^3 X } \right)\,,
\label{eq:p2}
\\
p &=& {\partial {\cal L} \over \partial \dot{X} }
- \nabla \left( {\partial {\cal L} \over \partial \nabla^2 X } \right) +
\nabla^2 \left( {\partial {\cal L} \over \partial \nabla^3 X } \right)\,.
\label{eq:p3}
\end{eqnarray}

The reason for adopting reparametrization covariant variables
should be clear at this stage already. Under a reparametrization,
$\xi \to f(\xi )$, the various momenta transform as scalar
densities of different weight. The only true scalar under
reparametrizations is $p$. If one uses $( \dot{X} , \ddot{X} ,
{\mathop{X}\limits^{\ldots}} )$ as configurations variables, the
intermediate steps of the calculations get cluttered by various
non-covariant terms, which disappear in the final results anyway.

One might be tempted to go further and employ derivatives of $X$ with respect to 
proper time as the configuration variables. However due care must then be taken with 
the fact that the boundary values develop a non-local dependence on the velocity \cite{guv1}.
The  high cost of this tradeoff nullifies the apparent advantage.

For two arbitrary phase space functions the Poisson bracket is given by
\begin{equation}
\{ f , g \} = {\partial f \over \partial \Pi } \cdot
{\partial g \over \partial \nabla^2 X } +
{\partial f \over \partial P }
\cdot {\partial g \over \partial \dot{X} }
+ {\partial f \over \partial p } \cdot
{\partial g \over \partial X } - ( f\leftrightarrow g )\,.
\label{eq:pb}
\end{equation}

The canonical Hamiltonian density is
\begin{equation}
{\cal H}_c = \Pi \cdot \nabla^3 X + P \cdot \nabla^2 X + p \cdot
\dot{X} - {\cal L}\,.
 \end{equation}

The invariance of the action under reparametrizations implies the
existence of constraints. According to the standard Dirac-Bergmann
theory, we will have $M$ primary constraints,  ${\cal
C}_\alpha$ ($\alpha = 1,2,\cdots, M $). Their conservations in
time may produce secondary constraints, whose conservation in time
may give tertiary constraints, until the process stops and no new
constraints are generated. The constraints are called first class
if they are in involution.  What is important for higher-derivative 
constrained systems is the
number of
primary constraints that remain first class at the end of the
Dirac-Bergmann algorithm. These are the generators of the
transformations that leave the action invariant. In our case, we
expect only one invariance, reparametrization invariance.

Time evolution is generated by  the extended Hamiltonian density
given by adding to the canonical Hamiltonian
 the primary constraints ${\cal C}_\alpha$,
\begin{equation}
{\cal H}_T = {\cal H}_c + \lambda^\alpha {\cal C}_\alpha\,.
 \end{equation}

\section{First curvature}

We restrict now our attention to models that depend only on the
first curvature $L = L (\kappa_1 )$. Our first task is to evaluate
the momenta. Using Eq. (\ref{eq:k1d}), we have
\begin{equation}
{\partial \kappa_1 \over \partial \nabla^2 X } =
 {1 \over \kappa_1 ( - \gamma)^2  } \; \nabla^2 X  =
{1 \over ( - \gamma) } \;  \hat{k}^i \; n_i = {1 \over ( - \gamma) }
\; \eta_1\,,
\end{equation}
where we have exploited  Eq. (\ref{eq:eta1}), so that the momenta $P$
associated with the velocities, as given by specializing  Eq.
(\ref{eq:p2}), using the Leibniz rule, are
\begin{equation}
P = {L_1 \over \sqrt{-\gamma}} \; \eta_1 \,, \label{eq:pg}
\end{equation}
where  $L_1 = dL / d\kappa_1 $. $P$ is  always normal to the
worldline, and in particular it is proportional to the first
F-S normal direction. Only the proportionality
coefficient is model dependent. This provides a simple geometrical
interpretation for $P$. In this class of models, the first
F-S normal is conjugate to the velocity $\dot{X}$.

For the momenta $p$ conjugate to the embedding functions,  first
note that
\begin{equation}
{\partial \kappa_1 \over \partial \dot{X} }
=  { 2 \kappa_1 \over (-\gamma)} \;  \dot{X}  =
{2 \kappa_1 \over \sqrt{-\gamma}} \;  X'
\,,
\label{eq:kv}
\end{equation}
so that, using the Leibniz rule and the expression
\begin{equation}
{\partial \sqrt{-\gamma}\over \partial \dot{X} }
= - {\dot{X} \over \sqrt{-\gamma}} = - X' \,.
\label{eq:sqrt}
\end{equation}
we obtain
\begin{equation}
{\partial {\cal L} \over \partial \dot{X} }
=  \left( 2 L_1 \, \kappa_1 - L \right) \; X'
\,,
\end{equation}
In order to evaluate the derivative
of $P $,  we can use the second F-S equation in Eqs. 
(\ref{eq:frenet}) to get
\begin{equation}
\nabla P = \sqrt{-\gamma} \; P{}' = ( L_1 \eta_1 ){}' =
 ( L_1 )' \; \eta_1 + L_1 \; \kappa_1 \;  X'   -
L_1 \; \kappa_2 \; \eta_2  \,.
\label{eq:naP}
\end{equation}
Inserting these  expressions into Eq. (\ref{eq:p3}) as 
specialized to the present case, the momentum conjugate to $X^\mu $
takes the form
\begin{equation}
p = \left( L_1 \; \kappa_1 - L \right) \; X'  - ( L_1 )'
\; \eta_{1}  +
L_1 \; \kappa_2 \; \eta_{2} \,.
\label{eq:ps}
\end{equation}
This expression coincides with the conserved linear momentum
associated with the Poincar\'e translational invariance of 
the action and obtained 
directly using
the Noether theorem (see {\it e.g} \cite{ACG2}). If the 
Lagrangian
depends on $\kappa_1$, there will be non-trivial normal 
component along the 
first two F-S normal directions;
the momentum possesses a non-vanishing spacelike component orthogonal to
the timelike particle trajectory. This is a manifestation of a
generic feature of higher-derivative theories. In general, there will be
also the possibility of tachyonic energy flow, since, in general, 
$M^2 = - p^2 $ may take
values of arbitrary sign, even if in the particle rest frame,
$ p\cdot X'  = - E $ may yield a positive energy $E$.

Note that in the special case linear in the curvature, $L =
\beta \; \kappa_1 $,  the tangential component vanishes. This is a
consequence of the scale invariance of this particular model.
Since this degenerate case has been studied extensively in the
literature (see {\it e.g.} \cite{Ply1,Ply2,KP, RR1,RR2}), 
henceforth we will restrict our attention to the
generic case.

According to the  Dirac-Bergmann theory of higher-derivative
theories we immediately identify the primary constraint, which involves
the highest momenta $P$,
\begin{equation}
{\cal C}_1 = P \cdot \dot{X} = 0\,.
\label{eq:c1}
\end{equation}
There are no other primary constraints, as one can verify by computing
the Hessian 
\begin{eqnarray}
H_{\mu\nu} &= &{ \partial^2 {\cal L} \over \partial \nabla^2 X^\mu
\partial \nabla^2 X^\nu } \nonumber \\
&= &{1 \over (-\gamma )^{3/2} \; \kappa_1 }
\left[ (\kappa_1 \; L_{11} - L_1 )\;  \eta_{\mu\, 1} \eta_{\nu\, 1}
+ L_1 \; ( X_\mu{}' X_\nu{}' 
+ \eta_{\mu\nu} ) \right]\,, 
\end{eqnarray}
where $L_{11} = d^2 L / d\kappa_1{}^2$. The only null eigenvector is $X'$.

Moreover note that squaring Eq. (\ref{eq:pg}), we have
\begin{equation}
(- \gamma) \; P^2 = L_1{}^2\,,
\end{equation}
which allows us to cast $\kappa_1$ with respect to canonical
variables in the scalar combination
$\gamma \; P^2$. Therefore, using $ P \cdot \nabla^2 X
= \sqrt{-\gamma} L_1 \; \kappa_1 $, the canonical 
Hamiltonian density can be written as
\begin{equation}
{\cal H}_c = p \cdot \dot{X} + \sqrt{-\gamma} \; V ( \gamma P^2 ) \,,
\label{eq:hc1}
\end{equation}
where the potential $V$ is defined by
\begin{equation}
 V ( \gamma \; P^2 ) =   L_1 \; \kappa_1 - L  \,,
\label{eq:pot}
\end{equation}
and it is understood to be an implicit function of
the phase space variables $(  \dot{X} , P ) $, in the
scalar combination $\gamma P^2$.
The potential has the form of a Hamiltonian
with respect to the ``velocities" $\kappa_1$, with momenta
$L_1$, and lagrangian $L$.

The Hamiltonian that generates the dynamics is given by
adding the constraint (\ref{eq:c1}) to the canonical
hamiltonian density  (\ref{eq:hc1}),
\begin{equation}
{\cal H}_T = {\cal H}_c + \lambda_1 \; {\cal C}_1\,,
\end{equation}
with $\lambda_1$ an arbitrary Lagrange multiplier.
Conservation in time of the primary constraint, using the Poisson bracket
(\ref{eq:pb}) 
gives
\begin{equation}
\nabla {\cal C}_1 = \{ {\cal H}_T , {\cal C}_1 \}
= - {\cal H}_c = 0\,.
\end{equation}
Therefore we have that the vanishing of the Hamiltonian,
as expected from reparametrization invariance,  follows
as a secondary constraint. There are no other constraints.
The phase space has  $4 (N+1)$ degrees of freedom. There are
two first class constraints, therefore the number of physical
degrees of freedom is $(1/2) [4(N+1) - 2 \times 2 ] = 2N $.

We can gain insight into the structure of the Hamilton equations
without needing to know the explicit form of the potential $V$.
The first Hamilton equation is, as expected, an identity
\begin{equation}
\dot{X} = \{ {\cal H}_T , X  \} = {\partial {\cal H}_T \over
\partial p } = \dot{X} \,.
\label{eq:h1}
\end{equation}
The second Hamilton equation identifies both the
Lagrange multiplier, $\lambda_1$, and the
higher momentum $P $. We have
\begin{equation}
\nabla^2 X = \{ {\cal H}_T , \dot{X} \} = {\partial
{\cal H}_T \over
\partial P  } = \lambda_1 \; \dot{X}
- 2   (-\gamma)^{3/2} \;  V^* \; 
P \,,
\label{eq:h2}
\end{equation}
where $V^*$ denotes derivative of $V$  with respect to its
argument.
Contraction of this expression with $\dot{X}$, and use
of the constraint
(\ref{eq:c1}) gives $\lambda_1 = 0$. This is clearly a
consequence of our choice for the configuration variables.
Using $\lambda_1 = 0$, we have that $P$ is proportional to
$\nabla^2 X$, or $\eta_1$, with Eq. (\ref{eq:eta1}).
Contraction with the first F-S normal gives
\begin{equation}
\kappa_1 = - 2  \sqrt{-\gamma} \;  V^* \; P \cdot \eta_1 \,,
\label{eq:h2a}
\end{equation}
which determines $P $ implicitly, so long as
$V^* \neq 0$. Indeed, as follows from the
definition of the potential, we know that
\begin{equation}
\kappa_1 = - 2 \; V^* \;L_1 \,,
\label{eq:vl}
\end{equation}
and so Eq. (\ref{eq:h2}) reproduces Eq. (\ref{eq:pg}).
Therefore, this Hamilton equation plays a role
analogous to the first Hamilton equation in the usual
case of a theory that depends at most on velocities,
reproducing as it does the definition of the momenta, $P$.

The third Hamilton equation determines the form of
the remaining momenta $p$. We have that
\begin{equation}
\nabla P = \{ {\cal H}_T , P \} = - {\partial
{\cal H}_T \over \partial
\dot{X} } =   -  p
+
\left( V + 2 \gamma \; P^2 \; V^* \right) \; X' \,.
\label{eq:h3}
\end{equation}
We need to compare Eq. (\ref{eq:h3}) with Eq. (\ref{eq:ps}).
However, using Eq. (\ref{eq:vl}) we have
\begin{equation}
V + 2 \gamma \; P^2 \; V^* = 2 L_1 \; \kappa_1 - L\,.
\label{eq:h3a}
\end{equation}
Thus, using this together with Eq. (\ref{eq:naP}), 
we reproduce the defining relation
(\ref{eq:ps}) for $p$.

Finally, the equations of motion take the form
\begin{equation}
\nabla p  = \{  {\cal H}_T , p  \}
= - {\partial {\cal H}_T \over \partial X } = 0\,.
\label{eq:h4}
\end{equation}
In the absence of external fields, an explicit dependence
of the Hamiltonian density on the embedding functions
would break translation
invariance.

We can also show that the following Bianchi identity holds,
\begin{equation}
\nabla p \cdot \dot{X} = 0\,.
\label{eq:bia}
\end{equation}
This follows from reparametrization invariance.
To see this in the
Hamiltonian context, note that
\begin{equation}
\nabla {\cal H}_c = \nabla p \cdot \dot{X} +
p \cdot \nabla^2 X - 2 (-\gamma )^{3/2} \; V^* \; P \cdot \nabla P
= \nabla p \cdot \dot{X}  =
0\,,
\end{equation}
where we have used Eqs. (\ref{eq:h2}), (\ref{eq:h3}).
The non-trivial part of the equations of motion is
given by the normal projection of  Eq. (\ref{eq:h4}),
\begin{equation}
\nabla p \cdot \eta_i = \sqrt{-\gamma} \; p{}' \cdot \eta_i = 0\,.
\label{eq:h4n}
\end{equation}
Indeed, as shown elsewhere \cite{ACG2},
we have in general
\begin{equation}
p{}' = {\cal E}^i \; \eta_i \,,
\label{eq:h4nn}
\end{equation}
where the right hand side represents the Euler-Lagrangian
derivative for this class of models.

Let us consider a specific model.
The simplest is a massive relativistic particle with a
correction quadratic in the geodesic curvature,
\begin{equation}
L (\kappa_1 )  = - m + {\alpha \over 2} \;  \kappa_1{}^2 \,,
\end{equation}
where $\alpha$ is a coupling constant with dimension
of length, and $m$ has dimension of inverse length.
(For different, but equivalent, approaches to the Hamiltonian
formulation of this particular model see Refs. \cite{Ply3,Nerse}.)
In the notation introduced in Eq. (\ref{eq:pot}), we have
that the potential is
\begin{equation}
V  = m - {1 \over 2 \alpha} \; \gamma \; P^2\,.
\end{equation}
It is therefore {\it linear} in the variable $\gamma P^2$.
The equations of motion (\ref{eq:h4n}) take the form
\begin{eqnarray}
p{}' \cdot \eta_1 &=& - \alpha \kappa_1{}'' +  {\alpha \over 2} \kappa_1^3
+
\alpha \kappa_1 \kappa_2^2 + m \kappa_1 = 0\,,
\nonumber \\
p{}' \cdot \eta_2 &=& 2 \kappa_2 \kappa_1{}' + \kappa_1 \kappa_2{}'=0\,,
\nonumber \\
p{}' \cdot \eta_3 &=& - \kappa_1 \kappa_2 \kappa_3 =0\,,
\nonumber
\end{eqnarray}
and it is easy to check that these equations coincide with 
the Euler-Lagrange equations for this
model (see {\it e.g.} \cite{ACG2}).
We note that only three projections survive. The projection along $\eta_3$ 
implies the vanishing of $\kappa_3$, the motion lives in a $2+1$ dimensional
space. Furthermore the second equation implies that $\kappa_2$ is a 
function of $\kappa_1$. With this fact, it is then clear that 
the first equation is always integrable \cite{ACG2, Nesto}.

\section{Second curvature}

We extend now our considerations
to models that depend on the second curvature
$L = L (\kappa_2 )$. To evaluate
the momenta we  need to know
how $\kappa_2$ depends on $( \nabla^3 X,
\nabla^2 X , \dot{X} )$. From Eq. (\ref{eq:k2d}), using Eq.
(\ref{eq:eta2}),
we obtain,
\begin{eqnarray}
{\partial \kappa_2 \over \partial \nabla^3 X }
&=& - {1 \over (-\gamma)^{3/2}\; \kappa_1 } \; \eta_2 \,,
\label{eq:kk2a}
\\
{\partial \kappa_2 \over \partial \nabla^2 X }
&=& {1 \over (-\gamma) } \left( {  \kappa_1{}'  \over 
\kappa_1^2 } \; \eta_2
- {  \kappa_2 \over \kappa_1 } \; \eta_1 \right) \,,
\label{eq:kk2b}
\\
{\partial \kappa_2 \over \partial \dot{X} }
&=& {1 \over \sqrt{-\gamma}}  \left( \kappa_1 \; \eta_2
+  \kappa_2 \; X' \right) \,.
\label{eq:kk2c}
\end{eqnarray}
It follows immediately that the highest momentum, as 
defined by Eq. (\ref{eq:p1}), is
\begin{equation}
\Pi = - {L_2 \over \sqrt{\nabla^2 X \cdot
\nabla^2 X  } } \; \eta_2\,,
\label{eq:p21}
\end{equation}
where $L_2 = d L / d\kappa_2 $.
It is proportional to the second
F-S normal, giving it a nice geometrical
interpretation. Up to a model dependent proportionality
factor the second F-S normal $\eta_2$ is conjugate to the 
acceleration $\nabla^2 X$.
 Notice the similarity in structure
with the highest momenta in the first curvature models
(\ref{eq:pg}).

For the momentum conjugate to $\dot{X}$, as given by Eq.
(\ref{eq:p2}), using Eq. (\ref{eq:kk2b}),
first we have that
\begin{equation}
{\partial {\cal L} \over \partial \nabla^2 X}
={ L_2 \over  \sqrt{-\gamma} } \left( {  \kappa_1{}'  
\over \kappa_1^2 } \;  \eta_2
- {  \kappa_2 \over \kappa_1 } \; \eta_1 \right) \,.
\end{equation}
Moreover, exploiting the F-S equations ({\ref{eq:frenet}),
we have that
\begin{equation}
\nabla \Pi = \sqrt{-\gamma}\;  \Pi{}' =
- {1 \over \sqrt{-\gamma}}
\left[ \left( {L_2 \over \kappa_1} \right)'
\; \eta_2 + { L_2 \kappa_2 \over \kappa_1 } \; \eta_1
- {L_2 \kappa_3 \over \kappa_1 }\;  \eta_3 \right]\,.
\end{equation}
Therefore we obtain, after some cancellations,
\begin{equation}
P  = {1 \over \sqrt{-\gamma}} \left( {  L_2{}' \over \kappa_1 } \; \eta_2
- {L_2 \kappa_3 \over \kappa_1 } \; \eta_3  \right)\,.
\label{eq:PP}
\end{equation}
The structure is completely different from what we found earlier
for the first curvature models. In particular, note that there is
no contribution along the first F-S normal, unless we
enlarge our considerations to models of the form
$ L (\kappa_1 , \kappa_2 )$. We will have more to say about this
possibility below.

Let us turn now to the momentum conjugate to $X$, $p$.
Using Eq. (\ref{eq:kk2c})
we have
\begin{equation}
{ \partial {\cal L} \over \partial \dot{X}} =
L_2 \; \kappa_1 \; \eta_2 + ( L_2 \; \kappa_2 - L ) \; X' \,.
\end{equation}
Moreover, exploiting the F-S equations (\ref{eq:frenet}),
we have
\begin{eqnarray}
\hspace{-0.5cm}\nabla P &=&
\nabla \left( {\partial {\cal L} \over
\partial \nabla^2 X }\right)
- \nabla^2 \left({\partial {\cal L} \over
\partial \nabla^3 X } \right) \nonumber \\
\hspace{-0.5cm}&=& {\kappa_2 \over \kappa_1} (L_2){}' \; \eta_1
+ \Bigg[  \left( { L_2{}' \over \kappa_1} \right)'
-  { L_2 \kappa_3{}^2 \over \kappa_1} \Bigg] \; \eta_2
- \Bigg[ \left( { L_2 \kappa_3 \over \kappa_1} \right)'
+ { L_2{}' \kappa_3 \over \kappa_1} \Bigg]\; \eta_3
\nonumber \\
&& + {L_2 \kappa_3 \kappa_4 \over \kappa_1 } \,\eta_4 \,.
\label{eq:napa}
\end{eqnarray}
so that inserting in Eq. (\ref{eq:p3}), we find
\begin{eqnarray}
\hspace{-0.5cm}p = (\kappa_2 \; L_2 - L )\; X'
&-&  {\kappa_2 \over \kappa_1} (L_2 ){}' \; \eta_1
+ \Bigg[  - \left( { L_2{}' \over \kappa_1} \right)'
+   { L_2 \over \kappa_1} ( \kappa_1^2  + \kappa_3^2 )\Bigg] \; \eta_2
\nonumber \\
&+& \Bigg[ \left( { L_2 \kappa_3 \over \kappa_1} \right)'
+ { L_2{}' \kappa_3 \over \kappa_1} \Bigg] \; \eta_3
-  {L_2 \kappa_3 \kappa_4 \over \kappa_1 } \,\eta_4 \,.
\label{eq:p23}
\end{eqnarray}
This expression coincides with the one obtained with the
conserved linear momentum obtained in \cite{ACG2}, using 
Noether's theorem. We
emphasize how much the intermediate calculations
are simplified by the use of the F-S equations.
Now there are non vanishing contributions along the first four F-S normals.

Using $\Pi \cdot \nabla^3 X = \sqrt{-\gamma} \; L_2 \;
\kappa_2$,
the canonical Hamiltonian takes the form
\begin{equation}
{\cal H}_c =
P \cdot \nabla^2 X + p \cdot \dot{X} + \sqrt{-\gamma}
\; V ( \nabla^2 X \cdot \nabla^2 X \,\Pi^2 ) \,.
 \end{equation}
where we use the relation
\begin{equation}
\Pi^2 (\nabla^2 X \cdot \nabla^2 X ) = L_2{}^2\,,
\end{equation}
to express $\kappa_2 $ in terms of phase space variables, and
the potential is given by
\begin{equation}
V = L_2 \; \kappa_2 - L\,.
\label{eq:v22}
\end{equation}
Here we emphasize the similarity with the expression
for the canonical density obtained in the first-curvature
models, Eq. (\ref{eq:hc1}). Again, the case linear in 
$\kappa_2$ is degenerate, and we exclude it from consideration.
We refer the reader interested in this special case to
Ref. \cite{RR2}. In the special case of a three-dimensional
background, there is also the possibility of a dependance 
linear in the {\it signed } torsion. This special case is
treated in Ref. \cite{Ply4,Nes3}

We recognize immediately the primary constraints
\begin{eqnarray}
{\cal C}_1 = \Pi \cdot \nabla^2 X &=& 0\,, \\
{\cal C}_2 = \Pi \cdot \dot{X} &=& 0 \,.
\end{eqnarray}
There are no other primary constraints, as one can 
see from the Hessian,
\begin{eqnarray}
\hspace{-2cm}
H_{\mu\nu} &=& { \partial^2 {\cal L} 
\over \partial \nabla^3 X^\mu
\partial \nabla^3 X^\nu }  \nonumber \\
\hspace{-2cm}
&=& {1 \over (-\gamma )^{5/2}
\; \kappa_1{}^2
\; \kappa_2 }
\left[ (\kappa_2 \; L_{22} - L_2 ) \; \eta_{\mu\, 2} \eta_{\nu\, 2}
+ L_2 ( X_\mu{}' X_\nu{}' - \eta_{\mu\, 1} \eta_{\nu\, 1} +\eta_{\mu\nu} ) 
\right]\,, 
\end{eqnarray}
where $L_{22} = d^2 L/ d\kappa_2{}^2 $.
The only null eigenvectors are $X{}'$ and $\eta_1$.

Note that the primary constraints are in involution,
\begin{equation}
\{ {\cal C}_1 , {\cal C}_2 \} = - {\cal C}_2 \,.
\label{eq:12}
\end{equation}
As discussed at the end of Sect. 3, the number of primary constraints
that remains first class after all the constraints have been generate
corresponds to the number of invariances of the action. Since we have
only reparametrization invariance we expect that one of these 
primary constraints will be second class. The constraint analysis below
shows that it is ${\cal C}_2$.

The total Hamiltonian is simply obtained by adding the
two primary constraints to the canonical Hamiltonian
\begin{equation}
{\cal H}_T =
{\cal H}_c + \lambda_1 \; {\cal C}_1 + \lambda_2\;
{\cal C}_2 \,.
\end{equation}

For variety, let us consider the Hamilton equations
first. We have the  identities:
\begin{eqnarray}
\nabla^2 X &=& {\partial {\cal H}_T \over \partial P}
= \nabla^2 X \,, \\
\dot{X} &=& {\partial {\cal H}_T \over \partial p}
= \dot{X}\,.
\end{eqnarray}

The form of $\Pi$ and of the
Lagrange multipliers is determined by the following
Hamilton equation
\begin{equation}
\nabla^3 X = {\partial {\cal H}_T \over
\partial \Pi } = 2 \sqrt{-\gamma} \;
V_* \; \nabla^2 X \cdot \nabla^2 X  \; \Pi
+ \lambda_1 \; \nabla^2 X + \lambda_2 \; \dot{X}\,,
\label{eq:hhh}
\end{equation}
where $V_*$ denotes the derivative of $V$ with respect to its argument.
Using the constraints, we obtain
\begin{eqnarray}
\lambda_1 &=&  {\nabla^3 X \cdot \nabla^2 X \over
\nabla^2 X \cdot \nabla^2 X } = { \nabla \kappa_1
\over \kappa_1 }\,, \label{eq:l1} \\
\lambda_2 &=&  {\nabla^3 X \cdot \dot{X} \over
\dot{X} \cdot \dot{X} } = (-\gamma)^{-1}
\nabla^2 X \cdot \nabla^2 X = (-\gamma ) \; \kappa_1{}^2 \,. \label{eq:l2}
\end{eqnarray}
Both multipliers are completely determined.
To obtain the form of $\Pi$, first note that using Eqs. 
(\ref{eq:eta2}), (\ref{eq:eta2a}) and the form of the multipliers, we
have that (\ref{eq:hhh}) implies that $\Pi$ is proportional 
to $\eta_2$. To obtain the proportionality factor, 
contraction of Eq. 
(\ref{eq:hhh}) with the second F-S
normal gives
\begin{equation}
- \kappa_2 = 2
V_* \; \sqrt{ \nabla^2 X \cdot \nabla^2 X } \; \Pi \cdot \eta_2\,,
\label{eq:pih}
\end{equation}
which determines $\Pi$ as long as $ V_* \neq 0$. Indeed
from the definition of the potential, we have
\begin{equation}
\kappa_2 =  2 V_* \; L_2\,,
\end{equation}
in parallel to expression (\ref{eq:vl}) for first curvature models,
so that Eq. (\ref{eq:hhh}) reproduces  Eq. (\ref{eq:p21}) for $\Pi$.
The next Hamilton equation determines the form of $P$,
\begin{equation}
\nabla \Pi = - {\partial {\cal H}_T \over \partial \nabla^2  X }
=  - P - 2 \sqrt{-\gamma} \; V_* \Pi^2 \nabla^2 X
- \lambda_1 \; \Pi \,.
\end{equation}
To see that it reproduces Eq. (\ref{eq:PP}) one
needs to use the expressions (\ref{eq:p21}) and
(\ref{eq:l1}) for $\Pi$ and $\lambda_1$, respectively.

The form of $p$ as given by Eq. (\ref{eq:p23}) is obtained from
\begin{equation}
\nabla P = - {\partial {\cal H}_T \over \partial \dot{X} }
=  - p +  V \; X' - \lambda_2 \; \Pi \,,
\end{equation}
as one can verify using Eq. (\ref{eq:p21}) for $\Pi$,
Eq. (\ref{eq:napa}) for $\nabla P$, Eq. (\ref{eq:v22}) for
$V$, and Eq. (\ref{eq:l2}) for $\lambda_2$.

Finally the equations of motion are given
\begin{equation}
\nabla p = - {\partial {\cal H}_T \over \partial X}
= 0 \,.
\end{equation}
As it was the case in first curvature models, we have
the Bianchi identity $\nabla p \cdot \dot{X} = 0$, and
the non-trivial part of the equations of motion is
given by the normal projections $\nabla p \cdot \eta_i = 0$.

We examine now the conservation in time of the
primary constraints.
We have that
\begin{eqnarray}
\nabla {\cal C}_1 &=& \{ {\cal H}_T , {\cal C}_1 \}
= - P \cdot \nabla^2 X + \lambda_2 {\cal C}_2\,,
\\
\nabla {\cal C}_2 &=& \{ {\cal H}_T , {\cal C}_2 \}
= - P \cdot \dot{X} - \lambda_1 {\cal C}_2\,.
\end{eqnarray}
Therefore we pick up  the obvious secondary constraints
\begin{eqnarray}
{\cal C}_3 = P \cdot \nabla^2 X &=& 0 \,, \\
{\cal C}_4 = P \cdot \dot{X} &=& 0 \,.\end{eqnarray}
These four constraints are in
involution; in addition to the bracket (\ref{eq:12}), we
have
\begin{eqnarray}
\{ {\cal C}_3 , {\cal C}_4 \} &=&
{\cal C}_3\,, \nonumber \\
\{ {\cal C}_3 , {\cal C}_1 \} &=&
- {\cal C}_3\,, \nonumber \\
\{ {\cal C}_2 , {\cal C}_3 \} &=&
{\cal C}_4 - {\cal C}_1  \,, \\
\{ {\cal C}_1 , {\cal C}_4 \} &=& 0 \,,\nonumber \\
\{ {\cal C}_4 , {\cal C}_2 \} &=&
{\cal C}_2 \,. \nonumber
\end{eqnarray}

The algebra generated by these four
constraints can be identified with $SO(3)$
times a line\footnote{We thank to
a referee for pointing this fact to us.}.
This can be interpreted  as a translation generated  by
the combination $J_0 =(1/2) ({\cal C}_1 + {\cal C}_4 )$, and a
rotation generated by $J_- ={\cal C}_2 , J_+ ={\cal C}_3 $
and $J_z=(1/2) ({\cal C}_1 - {\cal C}_4 )$. We can see
this by computing the change of the phase space variables
$\{ X,p;\dot{X},P; \ddot{X},\Pi \}$, caused by canonical
transformations with generators $J_0 ,\ldots,J_z$.

The conservation in time of the secondary constraint
${\cal C}_4$ implies the vanishing of the
canonical Hamiltonian, as 
expected from reparametrization invariance, but now
it is expressed as a sum of two constraints, i.e.,
the canonical hamiltonian is a sum of a secondary
and a tertiary constraint. We have
\begin{equation}
\nabla {\cal C}_4 = - p\cdot \dot{X} - \sqrt{-\gamma}\,V
+ {\cal C}_3  - \lambda_2 \; {\cal C}_2\,,
\end{equation}
which implies the tertiary constraint
\begin{equation}
{\cal C}_5 = p\cdot \dot{X} + \sqrt{-\gamma}\,V = 0\,,
\end{equation}
so that
\begin{equation}
{\cal H}_c = {\cal C}_3 + {\cal C}_5\,.
\end{equation}
Recall that for the first curvature models the vanishing
of the canonical Hamiltonian also followed from  the same constraint,
$P \cdot \dot{X} = 0 $. 

A further tertiary constraint arises from the
conservation in time of ${\cal C}_3$,
\begin{equation}
\!\!\!\!\!\!\!\!\nabla {\cal C}_3 = \{ H_T , {\cal C}_3 \} \approx 
\{ {\cal C}_5 , {\cal C}_3 \} =
 - p \cdot \nabla^2 X + 2 \sqrt{-\gamma} \;
V_* \; \nabla^2 X \cdot \nabla^2 X  \;  \Pi \cdot P  
 \,,
\end{equation}
which implies 
\begin{equation}
{\cal C}_6 = p \cdot \nabla^2 X - 2 \sqrt{-\gamma}
\; V_* \; \nabla^2 X \cdot \nabla^2 X \; \Pi \cdot P = 0\,.
\end{equation}

The constraint algebra is readily computed. We
have
\begin{eqnarray}
\{ {\cal C}_1 , {\cal C}_5 \} &=& 0 \,, 
\nonumber \\
\{ {\cal C}_2 , {\cal C}_5 \} &=& 0 \,, 
\nonumber \\
\{ {\cal C}_3 , {\cal C}_5 \} &=& {\cal C}_6 \,, \nonumber \\
\{ {\cal C}_4 , {\cal C}_5 \} &=& {\cal C}_5 \,, \\ 
\{ {\cal C}_1 , {\cal C}_6 \} &=& {\cal C}_6 \,, \nonumber \\
\{ {\cal C}_2 , {\cal C}_6 \} &\approx&
\sqrt{-\gamma} \; \left(  
2  \; V_*  \; \nabla^2 X \cdot \nabla^2 X  \; \Pi^2  - V\right) \,, 
\nonumber \\
\{ {\cal C}_3 , {\cal C}_6 \} &=& 2 \sqrt{-\gamma}
\; \nabla^2 X \cdot \nabla^2 X \; \left[
V_* P^2 + 2 V_{**} \nabla^2 X \cdot \nabla^2 X 
( \Pi \cdot P )^2 \right]\,, \nonumber \\
\{ {\cal C}_4 , {\cal C}_6 \} &=& 0 \,, \nonumber \\
\{ {\cal C}_5 , {\cal C}_6 \} &=& 4 \sqrt{-\gamma} V_* \nabla^2 X 
\cdot \nabla^2 X \Pi \cdot p\,, \nonumber 
\end{eqnarray}
where $V_{**}$ denotes the second derivative of $V$ with respect
to is argument. We emphasise that, to our knowledge, this is
the first time that
this algebra has been written down. The phase space has $6(N+1)$
degrees of freedom, there are 2 first class constraints, ${\cal C}_1$
and ${\cal C}_4$, and 4 second class constraints. Therefore the
number of physical degrees of freedom is $(1/2) [ 6 (N+1) - 2 \times 2
- 4 ] = 3N - 1 $.

Requiring stationarity of ${\cal C}_5$ we obtain nothing new.
As advertised earlier, the primary constraint ${\cal C}_2$ becomes second
class.
Therefore,  the conservation in time of ${\cal C}_6$ gives
\begin{eqnarray}
\nabla {\cal C}_6  &=& \{ {\cal H}_c , {\cal C}_6 \} \approx
\{ {\cal C}_3 , {\cal C}_6 \} +
\{ {\cal C}_5 , {\cal C}_6 \}
+ \lambda_2
\{ {\cal C}_2 , {\cal C}_6 \} \nonumber \\
&\approx&
2 \sqrt{-\gamma} \;
(\nabla^2 X \cdot \nabla^2 X ) \left[\,
 2 V_{**} \;
\nabla^2 X \cdot \nabla^2 X \; (\Pi \cdot P)^2
+ V_* \; P^2 \right.\nonumber \\
&& \left.+ 2 V_* \; \Pi \cdot p \,\right]
 + \sqrt{-\gamma} \; \lambda_2 \; \left[ \,
2 V_*  \; \nabla^2 X \cdot \nabla^2 X \;
\Pi^2  - V \,  \right] \approx 0 \,.
\end{eqnarray}
When the equations of motion are satisfied,
the value of the Lagrange multiplier
$\lambda_2$ coincides with the value obtained in Eq. (\ref{eq:l2}).

Let us specialize our considerations to the model
quadratic in $\kappa_2$,
\begin{equation}
L = {1 \over 2} \; \kappa_2{}^2\,.
\end{equation}
The canonical Hamiltonian density takes then the form
\begin{equation}
{\cal H}_c =
P \cdot \nabla^2 X + p \cdot \dot{X} + {\sqrt{-\gamma}
\over 2 }  (\nabla^2 X \cdot \nabla^2 X ) \Pi^2 \,.
 \end{equation}
Note that there are some simplifications in the brackets,
since $V_{**}$ vanishes. Despite this, the system becomes
second-class constrained.

The equations of motion for this model are
\begin{eqnarray}
p{}' \cdot \eta_1  &=&  2 \kappa_2 \left( {\kappa_2 {}' \over \kappa_1}
\right)'
+  {\kappa_2{}' \over \kappa_1} \kappa_2{}'
 -  {\kappa_2^2 \over \kappa_1} ( {3\over 2} \kappa_1^2
+ \kappa_3^2 ) = 0\,,
\nonumber \\
p{}' \cdot \eta_2 &=& \left( { \kappa_2{}' \over \kappa_1} \right)''
 - \left( { \kappa_2{}' \over \kappa_1} \right) (\kappa_2^2
+ \kappa_3^2 )
- ( \kappa_2  \kappa_1 )' - 2 \left( { \kappa_2 \kappa_3^2 \over \kappa_1}
\right)' \nonumber \\
&+&  \left( { \kappa_2 \kappa_3 \over \kappa_1} \right) \kappa_3{}' = 0\,,
\nonumber \\
p{}' \cdot \eta_3 &=&  -  3 \kappa_2{}'' {\kappa_3 \over \kappa_1}
-4 \kappa_2' \left({\kappa_3 \over \kappa_1}\right)'
+ \kappa_2{}' {\kappa_3{}' \over \kappa_1}
- \kappa_2 \left({\kappa_3 \over \kappa_1}\right)'' \nonumber \\
&+& { \kappa_2 \kappa_3 \over \kappa_1} (\kappa_1^2 + \kappa_3^2
+ \kappa_4^2 ) = 0\,,
\\
p{}' \cdot \eta_4  &=& 3 { \kappa_2{}' \over  \kappa_1 } \kappa_3 \kappa_4
+ 2 \kappa_2 \kappa_4 \left( {\kappa_3 \over \kappa_1 } \right)'
+ { \kappa_2  \over \kappa_1 } \kappa_3 \kappa_4{}' = 0\,, \nonumber \\
p{}' \cdot \eta_5 &=& - { \kappa_2 \kappa_3 \kappa_4 \kappa_5 \over
\kappa_1 } = 0\,. \nonumber
\end{eqnarray}
It is again easy to check that these equations coincide with
the Euler-Lagrange equations for this
model \cite{ACG2}.
We note that this time five projections survive. The projection
along $\eta_5$ implies the vanishing of $\kappa_5$, so that the
motion lives in a $4+1$ dimensional
space. In a $2+1$ dimensional space, with $\kappa_3=0=\kappa_4$, it has been
shown in \cite{ACG2} that this system of equations is integrable.

In this section, we have considered models that depend
only on $\kappa_2$. However, it is easy to see that our
considerations can be extended to the more
general possibility $L = L (\kappa_1 , \kappa_2 )$.
The primary constraints remain unchanged, but the canonical
Hamiltonian and the secondary constraints will include extra
terms that can be derived using the results of Sect. 4.
The general structure is the same.

\section{Conclusions}

We have considered a geometrical formulation of the Hamiltonian
analysis for higher-order relativistic particles. By using gauge
covariant variables and the F-S basis we have been able
to provide a complete treatment of geometrical models that depend
on the first two F-S curvatures. 

Our considerations can be extended to geometrical models that
depend on any F-S curvature, $\kappa_n$. We expect that the top momentum
conjugated to $\nabla^n X$
will be of the form
\begin{equation}
P_{n} = \pm {L_n \over \sqrt{\pm \nabla^n X \cdot \nabla^n X }} \eta_n\,.
\end{equation}
There will be $n$ primary constraints, of which $n-1$ will be second class,
with the remaining first class one corresponding to reparametrization 
invariance. 

The extension to a curved background, although problematic in the
degenerate case linear in the curvature \cite{Zoller}, 
is not  in the generic case. 
The momenta will change appropriately, picking up a term that depends
on the background Christoffel symbol.

Finally, the analysis presented in this paper should be of help in establishing
a geometrical approach to the Hamiltonian formulation of
higher order extended objects. Work along these lines is in progress.

\ack

R.C. and E.R. received financial support from CONACyT project
32187-E. J.G. received financial support from CONACyT project
32307-E as well as DGAPA project IN119799. E.R. would like to 
thank  Professor A. P. Balachandran of the Department of Physics
of Syracuse University for hospitality and acknowledges
support from a CONACyT post-doctoral fellowship.

\end{document}